\documentclass [12pt] {article}
\usepackage{amssymb,amsmath}
\newcommand{\cs}[3]{{{#3} \brace {#1 #2}}}

\newcommand{\dd}{\mbox{\rm d}}

\newcommand{\Gam}{\Gamma}

\newcommand{\p}{\partial}

\newcommand{\be}{\begin{equation}}
\newcommand{\bear}{\begin{eqnarray}}
\newcommand{\ear}{\end{eqnarray}}
\newcommand{\ee}{\end{equation}}
\newcommand{\lbl}{\label}
\newcommand{\bi}{\bibitem}
\newcommand{\ci}{\cite}
\newcommand{\pav}{Pav\v si\v c}

\newcommand{\hs}{\hspace}

\begin{document}

\

\hs{1.5cm}

\begin{center}
 {\bf \LARGE Path and Path Deviation Equations for $p$-branes}

\hs{2mm}

 M.Pav\v si\v c{\footnote{Jo\v zef Stefen Institute, Ljubljana, Slovenia \\
 email: matej.pavsic@ijs.si}}
 and  M.E.Kahil{\footnote{October University For Modern Sciences and Arts ,
  Giza, Egypt}} \footnote {The American University in Cairo, New Cairo, Egypt \\
email:  kahil@aucegypt.edu}

\hs{6mm}

{\bf Abstract}
\end{center}

\footnotesize

Path and path deviation equations for neutral, charged,
spinning and spinning charged test particles,  using a
modified Bazanski Lagrangian, are derived. We 
extend this approach to strings and branes.
We show how the Bazanski Lagrangian for charged point particles and
charged branes arises \`a la
Kaluza-Klein from the Bazanski Lagrangian in 5-dimensions.

\normalsize
\baselineskip .6cm

\section{Introduction}

Theories of relativistic extended objects, called branes,
have become one of the most promising branches of research in
theoretical physics (for a review see \ci{Duff}). They are
a natural generalization of the concept of point particle.
Branes occur at microscopic scale as a part of string/M-theory,
or as fundamental objects of their own. On the macroscopic scale,
a 3-brane living in a higher dimensional embedding space,
can describe the entire universe. For a review of many aspects
of embedding and the `brane world' scenarios see\,\ci{PavsicTapia}.

A starting point of a brane theory is the Dirac-Nambu-Goto action,
which gives the brane equations of motion---a generalization of the
point particle geodesic equation. As in the case of the point particle, the
embedding space in which the brane moves can be curved. It is important
to understand how a brane moves in such a classical background space,
and how it deviates from the motion of a nearby brane. A pioneering work
in that direction has been done by Roberts\,\ci{RobertsString,RobertsBrane}.

Geodesic deviation equations for the point particle
has been extensively studied in the literature. This is important,
because from studying the free fall of two nearby
objects we obtain the information on curvature of spacetime by using
geodesic deviation equations. For instance, Ellis and Van Elst\,\ci{Ellis}
used such equations for studying the structure of cosmological models. Recently,
Wanas and Bakry \cite{WanasBakry} utilized geodesic deviation equations
to examine the stability of some celestial objects. Also, there are some
attempts by Roberts \cite{Roberts99} who quantized  geodesic and geodesic
deviation equations. Bazanski\,\ci{Bazanski} discovered a very powerful
method to obtain the geodesic and geodesic deviation equations from one
single Lagrangian. Some authors, have
 applied this approach for examining path equations in different geometries
 than the Riemannian \cite{WanasKahil},  testing the effect of extra force
 on dynamical motion of spiral galaxies due to dark matter \cite{KahilHarko}
 and expressing the required paths of polyvectors as defined in curved
 Clifford spaces \cite{Kahil2007}.

The aim of this work is to derive the equations of minimal surface
and minimal surface deviation by using a generalized version of the Bazanski
Lagrangian for $p$-branes.
In section 2 we review how this works for the point particle
($p=1$), and explain how  Bazanski Lagrangian\,\ci{Bazanski}
can be extended to include not only neutral point particles but also charged,
spinning and spinning charged particles. In
section 3   we  present the relevant equations of
minimal surface and minimal surface deviation for $p$-branes, and the case of
spinning and
charged $p$-branes as well. In section 4 we show how the Bazanski Lagrangian
for charged point particles and charged $p$-branes follow \` a la Kaluza-Klein
from the Bazanski Lagrangian in 5-dimensions.
 Finally, we give some concluding remarks on how
this work could be extended.

\section{The Bazanski Approach  for the Point Particle}
 Geodesic and geodesic deviation equations for a relativistic point particle
 can be obtained simultaneously
 by using the Bazanski Lagrangian \cite{Bazanski}:
\begin{equation}
L= g_{\alpha \beta} u^{\alpha} \frac{D \psi^{\beta}}{Ds} .
\lbl{1}
\end{equation}
Here $u^\alpha \equiv d X^\alpha/d s$, where $X^\alpha$ are
the particle's coordinates, and $s$ the proper time, whereas
$\psi^\beta$ is the $s$-dependent deviation vector, associated
with a one parameter
family of geodesics $X^\mu (s,\epsilon)$ according to\,\ci{Bazanski}
\be
    \psi^\mu = \epsilon \frac{\p X^\mu}{\p \epsilon}\Biggl\vert_{\epsilon=0} .
\lbl{1a}
\ee    
Performing the variation of the action
\be I[X^\alpha, \psi^\alpha] =\int ds \, L
\lbl{1b}
\ee
 with respect to
the deviation vector ${\psi^{\rho}}$, we obtain the geodesic equation:
\begin{equation}
\frac{du^{\alpha}}{ds} +\cs{\mu}{\nu}{\alpha}u^{\mu}u^{\nu}=0
\lbl{2}
\end{equation}
If we vary the action (\ref{1b})
respect to $X^{\rho}$, then we obtain the geodesic deviation
equation:
\begin{equation}
\frac{D^2\psi^{\alpha}}{Ds^{2}} =
{R^{\alpha}}_{\beta \gamma \delta} \psi^{\gamma} u^{\beta}u^{\delta}
\lbl{3}
\end{equation}

The above Lagrangian can be generalized\,\ci{Kahil2006} to include the coupling
of a charged particle with the electromagnetic field:
\begin{equation}
L = g_{\alpha \beta} u^{\alpha}{\frac{D \psi^{\beta}}{Ds}} +
\frac{e}{m} F_{\alpha \beta}u^{\alpha} \psi^{\beta} .
\lbl{4}
\end{equation}
Then, instead of eq.\,(\ref{2}), we obtain
\begin{equation}
\frac{du^{\alpha}}{ds} +\cs{\mu}{\nu}{\alpha}u^{\mu}u^{\nu}=
\frac{e}{m}{F^\mu}_\nu u^{\nu} .
\lbl{5}
\end{equation}

On the other hand, if we consider the Lagrangian
\begin{equation}
L = g_{\alpha \beta} u^{\alpha}{\frac{D \psi^{\beta}}{Ds}} +
\frac{1}{2m} R_{\alpha \beta \gamma \sigma}u^{\alpha}
\psi^{\beta}S^{\gamma \sigma} ,
\lbl{6}
\end{equation}
we obtain\,\ci{Kahil2006} the Papapetrou equation:
\begin{equation}
\frac{d u^{\alpha}}{d s}+\cs{\mu}{\nu}{\alpha}u^{\mu}u^{\nu}=
\frac{1}{2 m}
{R^\alpha}_{\mu \nu \rho} S^{\rho \nu} u^{\mu} ,
\lbl{7}
\end{equation}
which, together with $D S^{\mu \nu}/D s =0$, is the equation of motion
for a relativistic top moving in a gravitational field background.\
The latter equation for $S^{\mu \nu}$ means that our object does not
precess. In the Lagrangian (\ref{6}) we consider the spin tensor $S^{\mu \nu}$
as a fixed quantity that is not varied. Later we shall consider an
action in which also $S^{\mu \nu}$ is a dynamical variable.

Dixon equation for rotating charged objects
\begin{equation}
\frac{du^{\alpha}}{d s}+\cs{\mu}{\nu}{\alpha}u^{\mu}u^{\nu}=
\frac{e}{m}{F^\mu}_\nu u^{\nu}+\frac{1}{2m}
{R^\alpha}_{\mu \nu \rho} S^{\rho \nu} u^{\mu}
\lbl{8}
\end{equation}
follows from the Lagrangian that combines eqs.\,(\ref{4}) and (\ref{6}):
\be
L = g_{\alpha \beta} u^{\alpha}{\frac{D \psi^{\beta}}{Ds}} +
\frac{e}{m} F_{\alpha \beta}u^{\alpha} \psi^{\beta}
+ \frac{1}{2 m}{R^\alpha}_{\mu \nu \rho} S^{\nu \rho} u^{\mu} .
\lbl{7b}
\ee

Whilst the conventional Bazanski Lagrangian (\ref{1}) gives geodesic
and geodesic deviation equations, the trajectories of charged and
spinning objects
can be obtained from the modified Lagrangians (\ref{4}),
(\ref{6}). This has led Kahil \cite{Kahil2006} to suggest
the following general Lagrangian:
\begin{equation}
L = g_{\alpha \beta} u^{\alpha}{\frac{D \psi^{\beta}}{Ds}} +
f_{\beta}\psi^{\beta}
\lbl{9}
\end{equation}
where
$${ f_{\beta} =  a_{1} F_{\alpha \beta} u^{\beta} +
a_{2} R_{\alpha \beta \gamma \delta} S^{\gamma \delta} u^{\alpha}} ,$$
Here $a_{1}$ and $a_{2}$ are parameters that may take the values
 ${\frac{e}{m}}$ and ${\frac{1}{2m}}$ to be adjusted with the original
 Lorentz force equation and the  Papapetrou equation as well as the Dixon equation.

Let us now consider the deviation equations that correspond to Lagrangians
(\ref{4}), (\ref{6}), and (\ref{7b}).

{(  i) Charged deviation equation}:
\begin{equation}
\frac{D^{2}\psi^{\alpha}}{Ds^{2}}=
{R^\alpha}_{\mu \nu\rho} u^{\mu}u^{\nu}\psi^{\rho} +
\frac{e}{m}({F^\alpha}_\nu \frac{D \psi^{\nu}}{Ds}+{F^\alpha}_{\nu ;
\rho}u^{\nu}\psi^{\rho})
\lbl{10}
\end{equation}

{( ii) Rotating deviation equations (without precession)}
\begin{equation}
\frac{D^{2}\psi^{\alpha}}{Ds^{2}}= {R^\alpha}_{\mu \nu\rho} u^{\mu}u^{\nu}\psi^{\rho}
+ \frac{1}{2m}( {R^\alpha}_{\mu \nu \rho} S^{\nu \rho} \frac{D \psi^{\mu}}{Ds}+
{R^\alpha}_{\mu \nu \lambda}{S^{\nu \lambda}}_{; \rho}u^{\mu}\psi^{\rho}
+ {R^\alpha}_{\mu \nu \lambda; \rho }S^{\nu \lambda} u^{\mu} \psi^{\rho})
\lbl{11}
\end{equation}

{(iii) Rotating charged deviation equations}:
$$
\frac{D^{2}\psi^{\alpha}}{Ds^{2}}=
{R^\alpha}_{\mu \nu\rho} u^{\mu}u^{\nu}\psi^{\rho}
+\frac{e}{m}({F^\alpha}_\nu \frac{D \psi^{\nu}}{Ds}+{F^\alpha}_{\nu ;
\rho} u^{\nu}\psi^{\rho})
$$
\begin{equation}
~~~~~~~~~~~~~~
+
\frac{1}{2m}( {R^\alpha}_{\mu \nu \rho} S^{\nu \rho} \frac{D \psi^{\mu}}{Ds}+
{R^\alpha}_{\mu \nu \lambda}{S^{\nu \lambda}}_{; \rho}u^{\mu}\psi^{\rho}
+ {R^\alpha}_{\mu \nu \lambda; \rho }S^{\nu \lambda} u^{\mu} \psi^{\rho})
\lbl{13}
\end{equation}
The above deviation equations (\ref{10}),(\ref{11}), and (\ref{13}) can
also be derived directly from the equations of motion
(\ref{5}),(\ref{7}), and (\ref{8}), respectively. For example,
Nieto et al.\,\ci{Nieto} derived in such a manner the deviation equation
(\ref{11}) from eq.\,(\ref{7}) for two nearby tops.

Papapetrou \cite{Papapetrou} has formed an equation of a rotating object
which is able to
precess:
\begin{equation}
 \frac{D}{Ds}( m u^{\alpha} + u_{\beta}\frac{D S^{\alpha \beta}}{DS})=
 \frac{1}{2}
{R^\alpha}_{\mu \nu \rho} S^{\rho \nu} u^{\mu}
\lbl{11a}
\end{equation}

 Kahil has suggested the following Lagrangian \cite{Kahil2006}:
\begin{equation}
L= g_{\alpha \beta} ( m u^{\alpha}
+ u_{\beta}\frac{D S^{\alpha \beta}}{Ds}) \frac{D \psi^{\beta}}{Ds}
 + \frac{1}{2} R_{\alpha \beta \gamma \delta} S^{\gamma \delta} u^{\alpha} \psi^\beta
 \lbl{11b}
\end{equation}
which gives the Papapetrou equation (\ref{11a}), and the following
path deviation equation:
$$
\frac{D^{2}\psi^{\alpha}}{Ds^{2}}= \frac{1}{m}
{R^\alpha}_{\mu \nu\rho}u^{\mu}( m \, u^{\nu}
+ u_{\beta}\frac{D S^{\nu \beta}}{Ds})\Psi^{\rho}
- \frac{1}{m}g^{\alpha \sigma}g_{\nu \lambda}u_\beta 
 (\frac{D S^{\lambda \beta}}{Ds})_{; \sigma} \frac{D \psi^{\nu}}{Ds}
$$
\begin{equation}
~~~~~~~+ \frac{1}{2 m} \left ({R^\alpha}_{\mu \nu \rho} S^{\nu \rho} \frac{D \psi^{\mu}}{Ds}+
{R^\alpha}_{\mu \nu \lambda}{S^{\mu \lambda}}_{; \rho}u^{\nu}\psi^{\rho}
 + {R^\alpha}_{\mu \nu \lambda; \rho }S^{\nu \lambda} u^{\mu} \psi^{\rho}
 \right )
 \lbl{16}
\end{equation}

Let us now consider the following action
$$
I[X^\alpha,\psi^\alpha, S^{\alpha \beta},\psi^{\alpha \beta}]
=\int d s \left [ g_{\alpha \beta} ( m u^{\alpha}
+ u_{\rho}\frac{D S^{\alpha \rho}}{Ds}) \frac{D \psi^{\beta}}{Ds}
 + \frac{1}{2} R_{\alpha \beta \gamma \delta} S^{\gamma \delta} 
 u^{\alpha} \psi^\beta \right . \hs{4cm}$$
\be
 \hs{4cm} \left . + S_{\alpha \beta} \frac{D \psi^{\alpha \beta}}{D s} 
  - u^\rho (\frac{D S_{\alpha \rho}}{D s} u_\beta -
   \frac{D S_{\beta \rho}}{D s} u_\alpha) \psi^{\alpha \beta} \right ]
 \lbl{11c}
\end{equation}
which generalizes the action that corresponds to the Lagrangian (\ref{11b}).
The action (\ref{11c}), in addition to the particle position $X^\alpha$, and
the deviation vector $\psi^\alpha$, is a functional of the spin tensor
$S^{\alpha \beta}$ and the spin deviation tensor $\psi^{\alpha \beta}$.
The latter quantity is defined with respect to a 1-parameter
family of spin tensors $S^{\alpha \beta} (s, \epsilon)$ according to
\be
  \psi^{\alpha \beta} 
  =\epsilon \frac{\p S^{\alpha \beta}}{\p \epsilon}\Biggl\vert_{\epsilon=0}
\lbl{7a}
\ee

The variation of the action (\ref{11c})
with respect to $\psi^\alpha$ gives again the Papapetrou equation (\ref{11a}),
whereas the variation with respect $\psi^{\alpha \beta}$ gives
\be
  \frac{D S_{\alpha \beta}}{D s} + u^\rho (\frac{D S_{\alpha \rho}}{D s} u_\beta -
   \frac{D S_{\beta \rho}}{D s} u_\alpha) = 0 ,
\lbl{11d}
\ee
which is the equation of of motion for the spin derived by
Papapetrou\,\ci{Papapetrou}.

Introducing
\be
   p_\alpha = m u_\alpha + u^\beta \frac{D S_{\alpha \beta}}{Ds}
\lbl{11e}
\ee
we can write eq.\,(\ref{11d}) in the form
\be
   \frac{D S_{\alpha \beta}}{D s}  = -(p_\alpha u_\beta - p_\beta u_\alpha) .
\lbl{11f}
\ee   

The variation of the action (\ref{11c}) with respect to $X^\alpha$ and
$S^{\alpha \beta}$, gives a  coupled system of equations for the
deviation vector $\psi^\alpha$ and the deviation tensor $\psi^{\alpha \beta}$.
But, if we use the spin equations of motion (\ref{11d}) in the action
(\ref{11c}), then the terms containing $\psi^{\alpha \beta}$
become $S_{\alpha \beta} \frac{D \psi^{\alpha \beta}}{Ds}
+\frac{D S^{\alpha \beta}}{Ds} \psi^{\alpha \beta}$, which is equal
the total derivative $\frac{D}{Ds} (S^{\alpha \beta} \psi^{\alpha \beta})$.
Therefore, the extra terms have no influence on the equations of
motions for variables $X^\mu$ and $\psi^\alpha$, and we remain
with the system, described by the action (\ref{11a})-(\ref{16}).

\section{The Bazanski Approach for the Brane}
In this approach we are going to introduce the following Lagrangian
as a counterpart of the Bazanki Lagrangian for the geodesic
and geodesic deviation equations of the branes in curved embedding space:
 \begin{equation}
 L= \kappa \sqrt{f}f^{ab}g_{\mu \nu} \partial_{a} X^{\mu} D_{b} \Psi^{\nu}.
\lbl{17}
\end{equation}
Here $\kappa$ is the brane tension, $f^{ab}$ is the inverse of the induced
metric on the brane,
$f_{ab}\equiv \p_a X^\mu \p_b X^\nu g_{\mu \nu}$,
$f\equiv \sigma {\rm det} f_{ab}$, where $\sigma$ is plus or minus sign, so that
$f$ is positive regardles of the signature of $f_{ab}$, and
$\partial_{a} X^{\mu}$ is the vector
velocity on the brane. The brane sweeps a worldsheet which is parametrized
by coordinates $\xi^a$, its embedding functions being $X^\mu (\xi^a)$,
and $\p_a \equiv \p/\p \xi^a$.
$\Psi^{\nu}$ is the deviation vector field,
associated with a one parameter family of minimal worldsheets $X^\mu (\xi^a,\epsilon)$
according to
\be
    \Psi^\nu = \epsilon \frac{\p X^\mu}{\p \epsilon}\Biggl\vert_{\epsilon=0}.
\lbl{18}
\ee

A Lagrangian, similar
to (\ref{17}), but with the ordinary derivative $\p_b \Psi^\mu$,
instead of the covariant derivative
$D_b \Psi^\mu = \p_b \Psi^\mu + \Gam_{\rho \sigma}^\mu \Psi_\sigma
\p_a X^\rho$, is considered in ref.\,\ci{RobertsBrane}.

Applying the variation with respect to the {\it deviation vector} $\Psi^\nu$,
the Euler-Lagrange equation becomes
\be
  \frac{1}{\sqrt{f}}\,\p_c (\sqrt{f} f^{ac} \p_a X^\mu) +
   f^{ab} \Gamma_{\alpha \beta}^\mu \p_a X^\alpha \p_b X_\beta =0,
  \lbl{19}
  \ee
which can be written more compactly as
\begin{equation}
 D_{c} ( f^{ac} \partial_{c} X^{\mu}) =0.
\lbl{20}
\end{equation}
where $D_c \equiv D/D \xi^a$ is analogous to $D/Ds$, and denotes
the covariant derivative with respect to
the embedding metric $g_{\mu \nu}$ and the world sheet metric $f_{ab}$
(see eq.(\ref{d8}).

Let us now derive the corresponding Euler-Lagrange equation, obtained
by varying the action (\ref{17}) with respect to $X^\mu$:
\be
\partial _c \frac{{\partial L}}{{\partial \partial _c X^\alpha  }}\,\,
 - \,\,\frac{{\partial L}}{{\partial X^\alpha  }}\, = 0 \\
\lbl{d1}
\ee
Using
\be
 \frac{\partial }{{\partial X^\alpha  }}\left( {\sqrt f f^{ab} } \right)
 = \frac{1}{2}\sqrt f (f^{ef} f^{ab}  - f^{ae} f^{bf}
 - f^{af} f^{be} )\partial _e X^\rho  \,\partial _f X^\sigma  \,g_{\rho \sigma ,\alpha }
\lbl{d2}
\ee
\be
 \frac{{\partial (\sqrt f f^{ab} )}}{{\partial \partial _c X^\alpha  }}
 = \sqrt f (f^{cf} f^{ab}  - f^{ac} f^{bf}
 - f^{af} f^{bc} )\partial _f X^\sigma  g_{\alpha \sigma }
 \lbl{d3}
\ee
and
\be
 \partial _c \Gamma _{\alpha \sigma }^\nu   =
 \partial _\rho  \Gamma _{\alpha \sigma }^\nu  \partial _c X^\rho
\lbl{d4}
\ee
we obtain
\be
 D_c D^c \psi _\alpha   + \,\,R^\nu  _{\alpha \sigma \rho } \,g_{\mu \nu }
  \,\,\partial ^c X^\mu  \partial _c X^\rho  \psi ^\sigma  \hs{7cm}  \nonumber
 \ee
\be
\hs{6mm} + \,(1/\sqrt {f\,} )\partial _c \left[ \sqrt f (f^{cf} f^{ab}  - f^{ac} f^{bf}
 - f^{af} f^{bc} )\partial _f X^\sigma  g_{\alpha \sigma }
 \partial _a X^\mu  D_b \psi ^\nu  g_{\mu \nu } \right]  \nonumber
 \ee
 \be
\hs{2cm} - \frac{1}{2}(f^{ef} f^{ab}  - f^{ae} f^{bf}  - f^{af} f^{be} )\partial _e X^\rho  \,\partial _f X^\sigma  \,g_{\rho \sigma ,\alpha } \partial _a X^\mu  D_b \psi ^\nu  g_{\mu \nu }  = 0 \\
\lbl{d5}
\ee
where
\be
 \sqrt {f\,} D_c D^c \psi _\alpha   = \partial _c (\sqrt f D^c \psi _\alpha  ) - \sqrt {f\,} \Gamma _{\rho \alpha }^\lambda  D^c \psi _\lambda  \partial _c X^\rho
\lbl{d6}
\ee
Eq.(\ref{d5}) is the equation of motion for the deviation vector. We can rewrite it
in a more compact form by taking into account
\be
\partial _c \sqrt f  = \sqrt f \,\Gamma _{cd}^d ,  ~~~~~~
\partial _c f^{ab}  =  - f^{ae} \,\Gamma _{ec}^b  - f^{be} \,\Gamma _{ec}^a
\lbl{d7}
\ee
where $\Gamma _{ec}^b$ is the connection on the brane's world manifold, and
using the covariant derivative defined according to\footnote{We distinguish the
two kinds of connection by indices only. In this case,
such simplified notation does not lead to confusion.}
\be
D_a D_b \psi ^\mu = \partial _a D_b \psi ^\mu   - \Gamma _{ab}^c D_c \psi ^\mu
  + \Gamma _{\rho \sigma }^\mu  D_b \psi ^\sigma  \partial _a X^\rho .
\lbl{d8}
\ee
So we arrive at the following deviation vector equation:
\be
 D_c (D^c \psi _\alpha   + \partial ^c X_\alpha  \,\partial ^b X^\mu  D_b \psi _\mu
  - \partial ^b X_\alpha  \partial ^c X^\mu  D_b \psi _\mu
   - \partial ^b X_\alpha  \partial _b X^\mu  D^c \psi _\mu  ) \hs{2cm} \nonumber
\ee
\be
\hs{6cm}
   + \,\,{R^\nu}_{\alpha \sigma \rho } \,g_{\mu \nu }
   \,\,\partial ^c X^\mu  \partial _c X^\rho  \psi ^\sigma  \, = 0,
\lbl{d9}
\ee
which can be written compactly as
\be
D_c ({f^{cb}}_{\alpha \mu} D_b \psi^\mu ) +
R^\nu  _{\alpha \sigma \rho } \,g_{\mu \nu }
   \,\,\partial ^c X^\mu  \partial _c X^\rho  \psi ^\sigma  \, = 0 ,
\lbl{d10}
\ee
where
\be
  {f^{cb}}_{\alpha \mu}\equiv f^{cb} g_{\alpha \mu}   + (f^{cd} f^{ab}  - f^{ac} f^{bd}
 - f^{ad} f^{bc} )\partial_d X_\alpha \partial_a X_\mu .
 \lbl{d11}
\ee

Since the covariant derivative of the metric vanishes,
$D_c f^{ab} = 0$, eq.\,(\ref{d9}) or (\ref{d10}) can be written
in the form
\be
   (g_{\alpha \mu} - \p^a X_\alpha \p_a X_\mu ) D_c D^c \psi^\mu
   -2D_c D^a X_\alpha \,\p^c X_\mu D_a \psi^\mu
   +{R^\nu}_{\alpha \sigma \rho } \,g_{\mu \nu }
   \,\,\partial ^c X^\mu  \partial _c X^\rho  \psi ^\sigma  \, = 0 .
\lbl{d12}
\ee

{\it Special case: Point particle}

Let us now consider a special case, in which the dimensionality of the
branes world manifold is $n=1$, i.e., the case of a point particle.
Then eq. (\ref{d9}) becomes:
\be
f^{00} D_0 (D_0 \psi _\alpha   - \partial _0 X_\alpha  \partial _0 X^\mu
 f^{00} D_0 \psi _\mu  ) + \,\,{R^\nu}_{\alpha \sigma \rho } \,g_{\mu \nu }
  \,\,f^{00} \partial _0 X^\mu  \partial _0 X^\rho  \psi ^\sigma  \, = 0
\lbl{d13}
\ee
Now $f^{00}  = \frac{1}{{f_{00} }},~~f_{00}  = \partial _0 X^\mu
 \partial _0 X^\nu  g_{\mu \nu }  \equiv \dot X^\mu  \dot X^\nu  g_{\mu \nu }$.
So we have
\be
 \frac{D}{{D\tau }}[(g_{\alpha \beta }
 - \frac{{\dot X_\alpha  \,\dot X_\beta  }}
 {{\dot X^\mu  \dot X^\nu  g_{\mu \nu } }})\frac{D}{{D\tau }}\psi ^\beta  ]
 +{R^\nu}_{\alpha \sigma \rho } \,g_{\mu \nu } \dot X^\mu  \dot X^\rho
 \psi ^\sigma   = 0 .
\lbl{d14}
\ee
This is the point particle geodesic deviation equation derived by
Bazanski \ci{Bazanski}.

Our action is invariant under reparametrizations of the world manifold coordinates
$\xi^a$. So there is a freedom to choose a gauge. In the case of the point
particle, we can choose a gauge such that
\be
   \frac{{\dot X}^\mu}{\sqrt{{\dot X}^2}} \frac{D\psi_\mu}{D \tau} = constant .
\lbl{d15}
\ee
Then eq.\,(\ref{d14}) simplifies to
\be
  \frac{D^2 \psi_\alpha}{D \tau^2}
 +R^\nu  _{\alpha \sigma \rho } \,g_{\mu \nu } \dot X^\mu  \dot X^\rho
 \psi ^\sigma   = 0 .
\lbl{d16}
\ee
Such equation we had in section 2 where we used the gauge in which $\tau$
was equal to the proper time $s$.

\subsection{Charged and spinning branes}

In order to describe a {\it charged bran} moving in a gravitational and
electromagnetic background field, let us generalize the
Lagrangian (\ref{17}) as follows:
\begin{equation}
 L= \kappa \sqrt{f}f^{ab}g_{\mu \nu} \partial_{a} X^{\mu} D_{b} \Psi^{\nu}
  + e^{a} F_{\mu \nu}\Psi^{\mu}\partial_{a} X^{\nu},
\lbl{22}
\end{equation}
where $e^{a}$ is the charge current density on the brane\,\ci{BarutPavsic}.
Applying the variation with respect to the deviation vector, we obtain

\begin{equation}
 D_{c} (f^{ac} \partial_{c} X^{\mu})
 =\frac{e^{a}}{\kappa \sqrt{f}}\,  {F^\mu}_{\nu} \partial_{a} X^{\nu}.
\lbl{23}
\end{equation}

In the case of {\it spinning (rotating) brane}, the corresponding path equation
which is the counterpart of Papapetrou equation of spinning particles,
is obtained from the Lagrangian
\begin{equation}
 L= \kappa \sqrt{f}f^{ab}g_{\mu \nu} \partial_{a} X^{\mu} D_{b} \Psi^{\nu}
 + \frac{1}{2} f^{ab}  R_{\mu \nu \rho \sigma}
 \Psi^{\nu}\partial_{a} X^{\mu} S_b^{\rho \sigma}.
\lbl{24}
\end{equation}
Here $S_a^{\rho \sigma}$ is the spin density current on the brane,
such that the integration over the brane's world sheet surface
element $\dd \Sigma^a$ gives the total spin, $S^{\rho \sigma}$, of the
brane:
\be
     \int \dd \Sigma^a S_a^{\rho \sigma} = S^{\rho \sigma} .
\lbl{25}
\ee
An example of such rotating or spinning brane is considered in
ref.\,\ci{PavsicRigidBrane}, where the extrinsic curvature term is
added to the minimal surface action. The brane's spin then arises from the
extrinsic curvature.

The variation of (\ref{24}) with respect to the deviation vector on
the brane gives
  \begin{equation}
 D_a (f^{ab} \p_b X^\mu ) =
 \frac{1}{2 \kappa \sqrt{f}}\,  f^{ab} R^{\mu}_{~\nu\rho\sigma} \partial_{a} X^{\nu}
  S_b^{\sigma \rho}.
 \lbl{26}
 \end{equation}
The above equation of motion is a particular case of the equation
that was derived in ref.\,\ci{PavsicRigidBrane}
for the brane with extrinsic curvature.

The deviation equation for {\it the charged brane}  can be obtained by
taking the variation of (\ref{22}) with respect to $X^\mu$. So we obtain:
\begin{equation}
D_c (f^{cb \alpha\mu} D_b \psi_\mu )=
 R^{\alpha}_{~\mu \nu\rho}
\partial_{a}X^{\mu}\partial^{a}X^{\nu}\Psi^{\rho}
+\frac{e^{a}}{\kappa \sqrt{f}}({F^\alpha}_\nu D_{a} \Psi^{\nu}+{F^\alpha}_{\nu ;\rho}
\partial_{a}X^{\nu}\Psi^{\rho})
\lbl{27}
\end{equation}

Similarly, by taking the variation of (\ref{24}) with respect to $X^\mu$,
we obtain the deviation equation for spinning brane:
$$
D_c (f^{c b \alpha \mu} D_b \psi_\mu )
=R^{\alpha}_{~\mu \nu\rho}
\partial_{a}X^{\mu}\partial^{a} X^{\nu}\Psi^{\rho}
+ \frac{1}{2 \kappa \sqrt{f}}(R^{\alpha}_{~ \mu \nu \rho}
 S_{a}^{\nu \rho} D^a \Psi^{\nu}
$$
\begin{equation}
~~~~~~~~~~~~~~~~~~~~~ + R^{\alpha}_{~\mu \nu \lambda}
{S_{a}^{\mu \lambda}}_{; \rho}
\partial^{a} X^{\nu}\Psi^{\rho}
 + R^{\alpha}_{~\mu \nu \lambda; \rho }S_{a}^{\nu \lambda}
 \partial^{a} X^{\mu} \Psi^{\rho})
.
\lbl{28}
\end{equation}

\section{Kaluza-Klein Approach}

Instead of considering a charged object in 4-dimensions, we can consider
a neutral object in 5-dimensional curved spacetime with metric
$G_{MN}$, $M,N = 0,1,2,3,5$. In the case of a {\it point particle},
the Bazanski Lagrabgian reads
\be
      L = M \, G_{MN} U^M \, \frac{D \psi^N}{DS} .
\lbl{4.1}
\ee
Here, constant $M$ is the mass in 5-dimensions, $U^M\equiv dX^M/dS$,
$\psi^N$ the geodesic deviation vector, and
$D \psi^N/DS = d\psi^N/dS + {\hat \Gam}_{JK}^N \psi^J U^K$ is the
covariant derivative with respect to the 5-dimensional line element
$dS=(G_{MN} dX^M dX^N)^{1/2}$.

For the 5-dimensional metric tensor we will take the following
Ansatz:
\be
    G_{MN} =  \begin{pmatrix}
                g_{\mu \nu} + A_\mu A_\nu &   A_\nu\\
                         A_\mu          &       1\\
              \end{pmatrix}
\lbl{4.2}
\ee
This is a simplified version of a more general 5D metric that,
besides the 4D metric $g_{\mu \nu}$ and the vector field
$A_\mu$,
also contains the scalar field $\phi = G_{55}$.

Using the relations
\be
    \psi_5 = G_{5 N} \psi^N = \psi^5 + A_\nu \psi^\nu
\lbl{4.3}
\ee
\be
    U_5 = G_{5 N} U^N = U^5 + A_\nu \psi^\nu
\lbl{4.4}
\ee
the Lagrangian (\ref{4.1}) can be written in the form
\be
   L = M \left ( g_{\mu \nu} U^\mu \frac{D\psi^\nu}{DS}
      + U_5 A_{\nu} \frac{D\psi^\nu}{DS} + U_5 \frac{D\psi^5}{DS} \right )
\lbl{4.5}
\ee
Further, if we take into account\footnote{Recall that $A_\nu = G_{5 \nu}$,
i.e., some components of the metric.} $D A_\nu/DS = 0$, then we have
\be
    L = M \left ( g_{\mu \nu} U^\mu \frac{D\psi^\nu}{DS}
    + U_5 \frac{D\psi_5}{DS} \right )
\lbl{4.6}
\ee

The components of the 5-dimenional connection split according to
\bear
&&{\hat \Gam}_{\nu \lambda}^\mu = \Gam_{\nu \lambda}^\mu
+ \frac{1}{2} (A_\lambda F_{\nu}^{~\mu} A_\nu F{\lambda}^{~\mu} ), \lbl{4.6a}\\
&&{\hat \Gam}_{\mu \nu}^5 = \frac{1}{2} (\nabla_\mu A_\nu + \nabla_\nu A_\mu)
  - \frac{1}{2} A^\rho (A_\nu F_{\mu \rho} + A_\mu F_{\nu \rho}), \lbl{4.6b}\\
&&{\hat \Gam}_{5 \nu}^\mu = \frac{1}{2} F_{\nu}^{~ \mu}~, ~~~~~
{\hat \Gam}_{5 \mu}^5 = - \frac{1}{2} A^\nu F_{\mu \nu}~, ~~~~~
{\hat \Gam}_{55}^\mu = 0,~~~ {\hat \Gam}_{55}^5 = 0 .  \lbl{4.6c}
\ear

From
\be
    \frac{D \psi^\nu}{D S} = \frac{d\psi^\nu}{dS}
    + {\hat \Gam}_{JK}^\nu \psi^J U^K ,
\lbl{4.7}
\ee
\be
   \frac{D \psi_5}{DS} = \frac{d \psi_5}{dS} - {\hat \Gam}_{5 N}^M \psi_M U^N ,
\lbl{4.8}
\ee
and $\psi_M = G_{MJ} \psi^J$, i.e., $\psi_\mu = G_{\mu \rho} \psi^\rho
+ G_{\mu 5} \psi^5$, $\psi_5 = G_{55} \psi^5 + G_{5 \mu} \psi^\mu$, where
$G_{\mu \rho} = g_{\mu \rho} + A_\mu A_\rho$, we find
\be
    \frac{D \psi^\nu}{D S} = \frac{d\psi^\nu}{dS} + \Gam_{\rho \sigma}^\nu
    \psi^\rho U^\sigma + \frac{1}{2} U_5 {F_\rho}^\nu \psi^\rho
\lbl{4.9}
\ee
\be
    \frac{D \psi_5}{DS} = \frac{d \psi_5}{dS}
    - \frac{1}{2} F_\nu^{~\mu} U^\nu g_{\mu \rho} \psi^\rho
\lbl{4.10}
\ee
The Lagrangian (\ref{4.6}) thus becomes
\be
    L = M g_{\mu \nu} U^\mu \left (\frac{d \psi^\nu}{dS}
    + \Gam_{\rho \sigma}^\nu \psi^\rho U^\sigma  \right )
    + M U_5 \, F_{\rho \mu} \psi^\rho U^\mu
    + M U_5 \frac{d \psi_5}{dS},
\lbl{4.11}
\ee
and the action is
\be
    I = \int L dS = \int L \frac{dS}{ds} ds =
     \int ds \left [ m g_{\mu \nu} u^\mu \left (\frac{d \psi^\nu}{ds} +
     \Gam_{\rho \sigma}^\nu \psi^\rho u^\sigma  \right )
     + e\, F_{\rho \mu} u^\mu + e \frac{d \psi_5}{ds} \right ] .
\lbl{4.12}
\ee
where
\be
   e=M U_5 = M\frac{dX_5}{dS} = m \frac{dX_5}{ds}~,~~~
   ds =(g_{\mu \nu}dX^\mu d X^\nu)^{1/2}
\lbl{4.13}
\ee
\be
       m = M \frac{ds}{dS}~,~~~~~{\rm and} ~~~~~  u^\mu = \frac{dX^\mu}{ds} .
\lbl{4.14}
\ee

The Lagrangian in eq.\,(\ref{4.12}) is just like that in eq.\,(\ref{4}),
apart from the additional term $e\, d\psi_5/ds$. One must also bear in mind
that $e = M U_5 =  M(U^5 + A_\nu U^\nu)$, and that because $e$ depends
on $U^\nu = dX^\nu/dS$, it contributes to the variation of the action
with respect to $X^\nu$, so that besides the terms occurring in eq.(\ref{10})
we obtain the additional term
\be
     F^\alpha_{~\lambda} U^\lambda \left [ \frac{d \psi_5}{dS}
     + F_{\rho \mu} \psi^\rho U^\mu \right ]
\lbl{4.14a}
\ee
If we perform the variation
with respect to $X^5$, we obtain the deviation equation for the fifth
component of the deviation vector:
\be
      \frac{d}{dS} \left [ \frac{d \psi_5}{dS}+ F_{\rho \mu} \psi^\rho U^\mu
      \right ] = 0.
\lbl{4.14b}
\ee
The latter equations are in agreement with the equations derived
by Kerner et al.\,\ci{Kerner} directly from the geodesic deviation
equation in five dimensions. Kerner et al.\ci{Kerner} observed that
the expression in square brakets of eq.\,(\ref{4.14a}) satisfies
equation (\ref{4.14b}), and is thus a constant of motion. In particular,
if the latter constant of motion is zero, then we have a one-to-one
correspondence between the geodesic deviation equation in 5
dimensions, derived from the action (\ref{4.12}) and the
usual deviation equation (\ref{10}) in presence of the electromagnetic
field in 4 dimensions. Those results hold for a particular Ansatz
(\ref{4.2}) with $G_{55} = 1$.

In the case of a {\it p-brane}, the Lagrangian (\ref{17}) can be extended
to
\be
    L = \kappa \sqrt{f} f^{ab} G_{MN} \p_a X^M D_b \Psi^N ,
\lbl{4.15}
\ee
where $M,N = \mu, 5$. A similar procedure as for the point particle gives
\be
   L = \kappa \sqrt{f} f^{ab} \left [g_{\mu \nu} \p_a (\p_b \Psi^\nu
      + \Gam_{\rho \sigma}^\nu \Psi^\rho \p_b X^\sigma)
      + \p_s X_5 F_{\rho \mu} \Psi^\rho \p_b X^\mu
      + \p_a X_5 \p_b \Psi_5 \right ] ,
\lbl{4.16}
\ee
which can be written as
\be
    L = \kappa \sqrt{f} f^{ab} g_{\mu \nu} \p_a X^\mu
      (\p_b \Psi^\nu + \Gam_{\rho \sigma}^\nu \Psi^\rho \p_b X^\sigma)
      + e^a F_{\rho \mu} \Psi^\rho \p_a X^\mu + e^a \p_a \Psi_5 ,
\lbl{4.17}
\ee
where $e^a \equiv \kappa \sqrt{f} f^{ab} \p_b X_5$. We see that the
Lagrangian (\ref{22}) is embedded in the Lagrangian (\ref{4.17}) which comes
from five dimensions. Instead of five, we can consider more dimensions
of the embedding spacetime.

\section{Discussion and Conclusion}

Since branes are so important objects considered in the attempts to develop
a unified theory of fundamental interactions, quantum gravity and the
brane world cosmological models,
we feel that such tasks require a thorough knowledge of all aspects
of the brane theory, starting with the most basic ones such as the
classical equations of motion.
We have shown how the path equations of motion for branes and the
corresponding deviation equations can be obtained from a single
Lagrangian (\ref{17}), which is a generalization of the Bazanski
Lagrangian. If a physical system consists of many branes, then their
relative motion can be described in terms of the
deviation equations. We have considered the minimal surface
deviation equation (\ref{d10}), and the deviation equations for charged
and spinning (rotating) branes.

The Bazanski action in 5-dimensions can be split into the action for
a charged object (point particle or a brane) in 4-dimensions plus the
extra terms. The latter terms enable distinction between the
4-dimensional and the 5-dimensional or, in general, a higher dimensional, theory.
Deviation equations for charged point particles
and branes can thus be used in testing the presence of extra dimensions.

A possible extension of the $p$-brane path and path deviation equations
is an analogous set of equations in  the
brane configuration space\,\ci{PavsicBook}, a theoretical framework which
has a potential to explain a deeper geometric principle behind brane
theory.

\section*{Acknowlgements}

Magd Kahil, the second author, would like to thank all members of
Jo\v zef Stefen Institute for inviting him as a visiting scientist
to their institute from 29 June to 10 July 2010.
This work was supported by the Ministry of High education, Science and
Technology of Slovenia.

 \end{document}